\documentclass[11pt,tightenlines,onecolumn, aps, nofootinbib, superscriptaddress, showkeys]{revtex4-1}

\usepackage{latexsym}
\usepackage{amssymb}
\usepackage[dvips]{graphicx}
\usepackage{color}
\usepackage{graphicx}

\usepackage{amsmath, amsfonts, amssymb, mathrsfs}
\usepackage{amsthm}
\usepackage{tensor}
\usepackage{multirow}
\usepackage{bbm}
\usepackage[geometry]{ifsym}

\newcommand{\Dslash}{{D\kern-0.63em{/}}}

\usepackage{graphicx}
\usepackage{float}
\usepackage[font=small]{caption}

\begin{document}

\newcommand{\beq}{\begin{equation}}
\newcommand{\eeq}{\end{equation}}

\title{From weak interaction to gravity}
\author{M. Novello}

\affiliation{ Centro de Estudos Avan\c{c}ados de Cosmologia (CEAC/CBPF) \\
 Rua Dr. Xavier Sigaud, 150, CEP 22290-180, Rio de Janeiro, Brazil. }

\author{A.~E.~S. Hartmann}
\affiliation{Dipartamento di Scienza e Alta Tecnologia, Universit\`{a} degli Studi dell'{}Insubria,  \\
 Via Valleggio 11, 22100 Como, and INFN Sez di Milano, Italy.}

\date{\today}

\begin{abstract}
We follow an old suggestion made by Stueckelberg that there exists an intimate connection between weak interaction and gravity, symbolized by the relationship between the Fermi and Newton\rq s constants. We analyze the hypothesis that the effect of matter upon the metric that represents gravitational interaction in General Relativity is an effective one. This leads us to consider gravitation to be the result of the interaction of two neutral spinorial fields (g-neutrinos) $ \Psi_{g}$ and $ \Omega_{g}$ with all kinds of matter and energy. We present three examples with only one g-neutrino: two static and spherically symmetric configurations and a cosmological framework for an isotropic dynamical universe. Without self-interaction, the associated effective geometry is precisely the Schwarzschild metric. On the other hand, a self-interacting g-neutrino generates a new gravitational black-hole.

\end{abstract}



 \maketitle

\section{Preamble}

The theory of General Relativity (GR) is based on two distinct (although related) hypothesis:
\begin{itemize}
\item{Gravity may be interpreted as nothing but a modification of the geometry of the underlying space-time;}
\item{The metric satisfies a dynamical equation that controls the action of matter on the modifications of the geometry.}
\end{itemize}

Although the first hypothesis have already demonstrated its success in describing a large number of phenomena, the second hypothesis has a more restrict support, it has a less sound acceptance. Indeed there is a profuse number of modifications of the GR dynamics like in \cite{joras, aldrovandi, cognola, LOOP, MOD, WIST, novellobergliaffa,percacci} and references therein. Some deals with corrections coming from quantum world, some coming from the need to avoid cosmological singularity and there are others that have a deeper origin in others theories of physics that intends to describe the dynamics of the geometry as a consequence of other interactions due to non-gravitational process ( see \cite{novellovisservolovik} for a review). We intend to implement this last idea.

\section{Pre-Introduction}

In a private communication at Geneva University in the early 1970 professor Stueckelberg pointed out to one of us (MN) that there is a curious property relating the Fermi constant of weak interaction and the Newton\rq s constant of gravity. He started by noting that in the system of units where $\hbar = c = 1$ both $ g_{w}$ and $g_{N}$ have dimension length square ($L^2$). Indeed, their values in this system is given by $g_{N} = N \, Gev^{-2}$ and $g_{w} = F \, Gev^{-2}$ where $ N = F^{8} = 10^{-40}.$ This does not occurs for the others two interactions (strong and electromagnetic). This led to the intriguing question \cite{Sivaram,Onofrio}: should these two universal interactions have some kind of relationship? In other words, should not gravity be constructed with the weak currents making the substitution of Fermi coupling $ g_{w}$ for Newton\rq s constant $\kappa?$

After the appearance of the gauge theories and the so-called weak-electromagnetic unification \cite{Salam,Weinberg} this problem lost its appeal. However, it seems that a return to that ancient idea could bring a new insight on gravitational processes. We would like to start to follow this path in the present work.
We will explore a procedure that combines the main ideas of General Relativity with the conjecture that Fermi interaction is intimately related to gravity, beyond the simple remark on the constants made by Stueckelberg. The recent interest by analog models of gravity and the idea that gravity should be nothing but an emergent theory of a more deep process set new ideas and perspectives in this direction.

We note from the beginning that gravity is well described by Einstein\rq s idea to associate universal gravity to the modification of the geometry of space-time. Fermi weak interaction deals with spinors and currents of the type $ \overline{\Psi} \, \gamma_{\mu} \, (1 - \gamma_{5} ) \Psi$ in the Minkowski background. For low energy we could apply the ancient current-current Lagrangian \cite{sudarshanmarshak} to describe their interaction. High energies provokes the appearance of an intermediate vectorial boson. Thus, one faces the problem to deal with two distinct languages to describe these two interactions. At this point we will make an appeal to some current ideas that investigate the possibility of the geometry used in gravitational interaction should be nothing but an effective one (see \cite{Sakharov,Volovik, Visser}). We will follow this procedure.

\section{Introduction}

The hypothesis that dynamical processes can be alternatively described in terms of a modification of the geometry of space-time has grown powerfully in the last years. In the case of GR the leitmotiv is very natural due to the universality of the gravitational interaction. On the other hand in effective theories the metric emerges as a consequence of the dynamical evolution of non-gravitational interactions. This geometric way has been considered not only for the evolution of the associated waves \cite{Hadamard} but also to nonlinear theories \cite{novellobittencourt}. The idea starts by writing the metric in the form
\begin{equation}
 g_{\mu\nu} = \eta_{\mu\nu} - h_{\mu\nu}
 \label{d1}
 \end{equation}
 This should be understood as an exact decomposition, it is not an approximation. Then, from mathematics it follows that its contra-variant expression, the inverse metric $ g^{\mu\nu}$ is given by an infinite series
 \begin{equation}
 g^{\mu\nu} = \eta^{\mu\nu} + h^{\mu\nu} - h^{\mu}{}_{\alpha}\,  h^{\alpha\nu} + h^{\mu}{}_{\alpha}\,  h^{\alpha\beta} \, h_{\beta}{}^{\nu}  + ...
 \label{d2}
 \end{equation}

The crucial novelty is contained precisely in this infinite terms. Any substitution of the geometrical approach done by GR by means of a corresponding field theory that deals with such infinity terms encountered hard difficulties. In an alternative to the geometrical approach some authors  have examined the possibility to understand this theory in terms of a spin-2 field in the flat Minkowski background\footnote{Note that once the theory is valid for any coordinate system, the symbol $\gamma_{\mu\nu}$ is in general used to denote the Minkowski metric in an arbitrary coordinate system.} \cite{Feynman,Grischuck84}. 

The question we would like to consider concerns the condition under which this series is limited. This is contained in the following 

{\sc Lemma} (The closure relation): the necessary and sufficient condition for the metric tensor defined as
\begin{equation}
g_{\mu\nu} = \eta_{\mu\nu} - h_{\mu\nu}
\label{7maio1}
\end{equation}
to have an inverse with finite terms is given by the closure relation
 \begin{equation}
  h_{\mu\nu} \, h^{\nu}{}_{\lambda} = \alpha \, \eta_{\mu\lambda} + \beta \, h_{\mu\lambda}
 \label{2}
 \end{equation}

 This allows us to write the inverse metric tensor under the restricted form
 $$g^{\mu\nu} = a \,\eta^{\mu\nu} + b \, h^{\mu\nu}.$$
We have the relations
$$ a = \frac{1 - \beta}{1 - \alpha - \beta}$$
$$ b= \, \frac{1}{1 - \alpha - \beta}.$$

 This is the case for the effective geometries constructed in theories of  scalar and electromagnetic fields \cite{novellobittencourt}.

\section{The gravitational neutrinos $ \Psi_{g}$ and $\Omega_{g}$}

The hypothesis we would like to develop here is based on the existence of two fundamental massless spinorial fields, that we will call gravitational neutrino (g-neutrino), which are the true elementary structure of gravity \cite{STG06,STG07}. We assume that these two fields  $ \Psi_{g}$ and $\Omega_{g}$ generate an effective metric which is the one dealing in General Relativity. In other words, the proposal of GR that gravity deals with metric\rq s modification of space-time has a substratum that we identify with these g-neutrinos.

Although the present proposal deals with two fields $ \Psi_{g}$ and $ \Omega_{g},$ in order to simplify our presentation in the present paper we consider a limited framework and deal only with a single one $ \Psi_{g}.$ In a subsequent paper we will analyze the full theory.

The dynamics of $ \Psi_{g}$ is given by the Dirac equation
\begin{equation}
\gamma^{\mu} \,\nabla_{\mu} \,\Psi_{g}= 0\,,
\label{26maio1}
\end{equation}
where
\begin{equation}
 \nabla_{\mu} \,  = \partial_{\mu} \, -\, \Gamma_{\mu}
 \label{7maio2}
 \end{equation}
and the internal connection contains beyond the conventional Fock-Ivanenko term an additional one driven by a scalar field, that is
\begin{equation}
\Gamma_{\mu} = \Gamma_{\mu}^{FI} + U_{\mu}\,.
\label{5abril5}
\end{equation}
The origin of this term will be explained below. Note that in the case we use Euclidean coordinates, the gamma\rq s are constant and the Christoffel symbol vanishes. Then the Fock-Ivanenko connection vanishes, once we have

\begin{equation}
 \Gamma_{\mu}^{FI} = - \, \frac{1}{8} \, \left( \gamma^{\nu} \, \partial_{\mu} \, \gamma_{\nu} -
\partial_{\mu} \gamma_{\nu} \, \gamma^{\nu} - \Gamma^{\alpha}_{\mu\nu} \, (\gamma^{\nu} \, \gamma_{\alpha} - \gamma_{\alpha} \, \gamma^{\nu}) \right)\,.
\label{7maio1}
\end{equation}

Let us note that we use indistinctly the symbol $\partial_{\nu}$ or a simple comma ($,$) to represent simple derivative. In this case covariant derivative will accordingly be noted with a semi-comma $ ( ;).$ The free spinor $ \Omega_g$ satisfies the same equation (\ref{26maio1}).

The total interaction between $\Psi_g$ and $\Omega_g$ is given by
\begin{align}
S   =  S_w + S_H\,,
\end{align}
where $S_w$ is denoting the V--A weak current,
\begin{equation}
S_w = \int \sqrt{- \eta} \, g_{w}\, (\overline{\Psi}_g  \gamma_{\mu} (1- \gamma_{5}) \Omega_g) \,  (\overline{\Omega}_g  \gamma^{\mu} (1- \gamma_{5})\, \Psi_g),
\label{27abril5}
\end{equation}
and $S_H$ is the Heisenberg potential \cite{heisenberg,heisenberg2} with two spinors,
\begin{align}
S_H = \int \sqrt{-\eta} \, s\, \Bigl\{   (\overline{\Psi}_g \, \Omega_g) \, (\overline{\Omega}_g \,\Psi_g) + (\overline{\Psi}_g \, \gamma_{5} \Omega_g) \, (\overline{\Omega}_g \, \gamma_{5} \,\Psi_g)\Bigr\}. \label{Heisenberg-interaction}
\end{align}
The real parameter $s$ has dimension of length squared.

Just for completeness we remind the Fierz-Pauli-Kofink (FPK) identities that establishes a set of tensor relations concerning elements of the four-dimensional Clifford $\gamma$-algebra. For any element $Q$ of this algebra the FPK
relation states the validity of the following expressions
\begin{equation}
(\overline{\Psi} \,Q \gamma_{\lambda}\, \Psi)\, \gamma^{\lambda} \Psi  =
(\overline{\Psi} \,Q\, \Psi)\,  \Psi  -  (\overline{\Psi} \,Q \gamma_{5}\, \Psi)\,
\gamma_{5} \Psi\,, \protect\label{H5}
\end{equation}
for $Q$ equal to $I$, $\gamma^{\mu}$, $\gamma_{5}$ and $\gamma^{\mu}
\gamma_{5}$. As a consequence of this relation we obtain two
 important consequences:
\begin{itemize}
 \item{The norm of the currents $J_{\mu} = \overline{\Psi}_g \, \gamma_{\mu} \, \Psi_g\,$  and  $\,I_{\mu} = \overline{\Psi}_g  \gamma_{\mu} \, \gamma_{5} \,\Psi_g\,$ have the same strength and opposite sign.}
 \item{The vectors  $J_{\mu}$ and $I_{\mu}$ are orthogonal.}
\end{itemize}
Indeed, using the FPK relation we have

\begin{align}
J^{\alpha} \, \gamma_{\alpha} \,\gamma_{5} \, \Psi &=(\overline{\Psi} \, \gamma_{5}
\Psi)\, \Psi -\, (\overline{\Psi}\,\Psi) \, \gamma_{5}\, \Psi \,,
\label{7maio3} \\[2ex]
J^{\alpha} \, \gamma_{\alpha} \, \Psi &=(\overline{\Psi} \,\Psi)\, \Psi -\, (\overline{\Psi}\, \gamma_{5} \,\Psi) \,\gamma_{5} \, \Psi \,,
\label{7maio4}
\\[2ex]
I^{\alpha} \, \gamma_{\alpha} \,\gamma_{5} \, \Psi &=(\overline{\Psi} \, \gamma_{5}
\Psi)\,\gamma_{5}\, \Psi -\, (\overline{\Psi}\,\Psi) \, \Psi \,,
\label{7maio5}
\\[2ex]
 I^{\alpha} \, \gamma_{\alpha} \, \Psi &=(\overline{\Psi} \,\Psi)\,\gamma_{5}\, \Psi -\, (\overline{\Psi}\, \gamma_{5} \,\Psi) \, \Psi \,.
\label{7maio6}
\end{align}

From (\ref{7maio4}) and (\ref{7maio6}) it follows that the norm of $J_{\mu}$ and $ I_{\mu}$ are given by
\begin{equation}
J^{\mu}\, J_{\mu} =  A^{2} + B^{2} = -  I^{\mu} I_{\mu}\,,  \protect\label{H7}
\end{equation}
where $ A \equiv \overline{\Psi} \, \Psi$ and  $ B \equiv i \, \overline{\Psi} \gamma_{5} \, \Psi$\,. Moreover the four-vector currents are orthogonal,
\begin{equation}
I_{\mu} J^{\mu} = 0. \protect\label{H71}
\end{equation}
From these results we have that the current $J_{\mu}$ is a time-like vector and the axial current is space-like.

\section{The origin of the covariant derivative of the gamma\rq s}

The covariant derivative of a spinor is given by eq. (\ref{7maio2}). Note that we use indiscriminately either the symbol semi-comma either nabla, that is $ \Psi_{; \, \mu} \equiv \nabla_{\mu} \, \Psi.$ In general, one obtains the explicit form of the internal connection $ \Gamma_{\mu} $ by the Riemannian condition that the covariant derivative of the metric tensor must vanish. A sufficient condition for this is to suppose that the covariant derivative of the gamma\rq s with which we construct the metric vanishes too.
Indeed, the vanishing of the covariant derivative of the gamma\rq s imply $ g_{\alpha\beta}{}_{; \lambda} =0.$ We remind the form of the covariant derivative of the gamma\rq s
\begin{equation}
  \gamma_{\nu \, ; \, \mu} = \partial_{\mu} \gamma_{\nu} - \Gamma^{\alpha}_{\mu\nu} \, \gamma_{\alpha} + \gamma_{\nu} \, \Gamma_{\mu} - \Gamma_{\mu}\, \gamma_{\nu}\,.
  \label{7maio8}
  \end{equation}

The metric $ \gamma_{\mu\nu}$ is defined from the fundamental objects, by
$$ \gamma_{\mu\nu}= \frac{1}{2} \, (\gamma_{\mu} \, \gamma_{\nu} + \gamma_{\nu} \, \gamma_{\mu}), $$
which is a multiple of the identity of the Clifford algebra.

In the case we assume $ \gamma_{\nu\, ; \, \mu}= 0$ a rather simple calculation yields the Fock-Ivanenko connection eq. (\ref{7maio1}). However, the Riemannian condition allows a more general expression for the internal connection. Indeed, it is a direct calculation \cite{novello73} to show that the vanishing of the covariant derivative of the metric tensor is satisfied if the evolution of the gamma\rq s is provided by the condition

\begin{equation}
\gamma_{\mu \, ; \, \nu} = [U_{\nu}, \gamma_{\mu} ]
\label{5abril2}
\end{equation}
where $ U_{\mu} $  besides to be a vector is an arbitrary object from the associated Clifford algebra. The simplest cases that we deal with here are provided by the expression

\begin{equation}
U =  \varepsilon \, H\,,
\label{17maio2}
\end{equation}
which appeals to an external scalar field $ H.$  We write
\begin{equation}
U_{\mu} = \frac{1}{4} \, \gamma_{\mu} \, \gamma^{\alpha} \, U_{, \, \alpha}
\label{5abril3}
\end{equation}

where $ U_{,\alpha} = \partial_{\alpha} U.$ This form implies the Riemannian condition for the metric, that is,
$$ \gamma_{\mu\nu \, ; \, \lambda} = 0$$ and the factor $ 1/4 $ is just for latter convenience.

\vspace{.5cm}

In short, the structure that we are dealing here contains three elements:
\begin{itemize}
\item{A manifold endowed with a Clifford algebra represented by $ \gamma_{\mu}$;}
\item{ A scalar field H;}
\item{Two fundamental spinors $ \Psi_{g}$ and $\Omega_{g}$.}
\end{itemize}

\subsection{Dynamics of H and the covariant derivative of the gamma\rq s}

The Lagrangian $ L_{0}$ of the scalar field yields the action
\begin{equation}
 S =  \frac{1}{2}\, \int \sqrt{-det \, \gamma_{\alpha\beta}}  \, \, H_{\mu} \, H_{\nu} \, \gamma^{\mu\nu}
 \label{7maio12}
 \end{equation}
 where $ H_{\mu} = H_{, \, \mu}$ is the derivative of $ H.$  Let us add to the free Lagrangian $ L_{0}$ a Lagrange multiplier by setting
\begin{equation}
 L = L_{0} + \varepsilon^{\mu\nu} \left( \gamma_{\mu ; \nu} + 2 \, \gamma_{\mu\nu} \, \gamma^{\alpha} \, H_{\alpha} - 2 \, \gamma_{\nu} \, H_{\mu}\right)
 \label{7maio16}
 \end{equation}
where $ \varepsilon^{\mu\nu}$ is a member of the Clifford algebra. Varying the Lagrange multipler we obtain the form of the covariant derivative of the gamma\rq s in the equation (\ref{5abril2}, \ref{5abril3}) and set the variation of the field $ H$  as $\delta L_{0} / \delta H  + 2 \,I. $
Imposing that $ I $ vanishes, that is,
$$ I = - \,( \varepsilon \, \gamma^{\alpha} )_{; \, \alpha} + ( \varepsilon^{\mu\nu} \, \gamma_{\nu} )_{; \, \mu} = 0,$$
a possible solution is given by
$$ \varepsilon_{\mu\nu} = \gamma_{\mu} \, \gamma_{\nu}.$$
This procedure provides the form of the covariant derivative of the gamma\rq s and the dynamics of the scalar field $H$ that is obtained from the free part $ L_{0}$ uniquely, that is
$$ \Box \, H = 0.$$.

\section{Fermi interaction and Gravity}

The weak interaction deals with three kinds of neutrinos. In the gravitational process we deal with two fundamental distinct kind of massless spinorial fields. We analyze the consequences of assuming the hypothesis that $ \Psi_{g}$ and $\Omega_{g}$ interacts universally with all matter through the modification of the geometry of the space-time according to the main principle of General Relativity.

In agreement with the framework suggested by Stueckelberg\rq s remark we define the gravitational metric $ g_{\mu\nu}$ in terms of the vectors\footnote{Let us point out that one could introduce two more null vectors constructed with these two spinors, that are right-hand currents. In this case the two g-neutrinos may be used to define a tetrad. In the presente work and once the gravitational metric is just an effective one (that is, it does not have a dynamics by its own) we restrict the set of the possible geometries to $\Delta_\mu$ and $\Pi_\mu$.  If, in the future, observations of more general metrics appears indispensable, the way is open to such generalization.} $ \Delta_{\mu} $ and $ \Pi_{\mu}$ constructed as a combination of the weak current $ \overline{\Psi}_g \, \gamma_{\mu} \, (1 - \gamma_{5}) \, \Psi_g$ and $ \overline{\Omega}_g \, \gamma_{\mu} \, (1 - \gamma_{5}) \, \Omega_g\,,$ that is, we set
\begin{equation}
 g_{\mu\nu} = \eta_{\mu\nu} -  \kappa \, h_{\mu\nu}\,,
 \label{9maio1}
 \end{equation}
where
\begin{equation}
h_{\mu\nu} = \,\Delta_{\mu} \, \Delta_{\nu}  + \Pi_{\mu} \, \Pi_{\nu} + \varepsilon \, (\Delta_{\mu} \, \Pi_{\nu} + \Delta_{\nu} \, \Pi_{\mu} )\,.
\label{27abril1}
\end{equation}
Let us remind that this is an exact form. It is not an approximation. By dimensionality argument we set the null-vectors defined in terms of the vector and axial currents of the spinorial fields, that is, the vectors $ \Delta_{\mu}$ and  $\Pi_{\mu} $ of dimensionality $ L^{- 1}:$
\begin{equation}
\Delta_{\mu} =  ( J_{\mu} - I_{\mu} ) \left(\frac{g_{w}}{J^{2}}\right)^{1/4}\,,
\label{18abril4}
\end{equation}

\begin{equation}
\Pi_{\mu} =  ( j_{\mu} - i_{\mu} ) \left(\frac{g_{w}}{ j^{2}} \right)^{1/4}\,,
\label{27abril2}
\end{equation}
where the currents $j_{\mu}$ and $  i_{\mu} $ are defined  in the same way as for the case of  $\Psi_g$\,, that is
$j_{\mu} = \overline{\Omega}_g \, \gamma_{\mu} \, \Omega_g\,, \; i_{\mu} = \overline{\Omega}_g  \gamma_{\mu} \, \gamma_{5} \,\Omega_g \,,\;  j^2 = j_{\alpha} \, j^{\alpha} = j_{\alpha} \, j_{\beta} \, \eta^{\alpha\beta} $  and  $ J^2 = J^{\alpha} \, J_{\alpha}.$ In the expression of the effective metric (\ref{9maio1}) we use the analogy of the current-current interaction displayed in eq. (\ref{27abril5}) and substitute $ g_{w}$ by $ \kappa,$ according to Stueckelberg suggestion.

The closure relation restricts the value of $\varepsilon^{2} = 1.$ We then have
$$ h_{\mu\nu} \, h^{\nu}_{\lambda} = 2 \, \varepsilon \, \Delta_{\nu} \Pi^{\nu} \, h_{\mu\lambda}\,,$$
which allows to write the inverse contra-variant metric as
\begin{equation}
 g^{\mu\nu} = \eta^{\mu\nu} + \, \frac{\kappa}{(1 - 2 \varepsilon \kappa \, \Delta_{\alpha}  \Pi^{\alpha})} \, h^{\mu\nu}.
 \label{7maio20}
 \end{equation}
Let us pause for a while and retain the main ideas of the present proposal, that is:
\begin{enumerate}
\item{The geometry (spin-2 field) of General Relativity is an effective representation of two physical (spin-1/2) fields, called \textit{g-neutrinos} and denoted by $ \Psi_{g}$ and $\Omega_{g}$, living in the Minkowski spacetime;}
\item{These fields satisfy the linear Dirac equation of motion. The covariant derivative of these spinors is controlled by a scalar field $ H .$}
\item{$\Psi_{g}$ and $\Omega_{g}$ interact via Fermi process;}
\item{Both fields couple with all forms of matter in an universal way;}
\item{ This interaction can be described as a modification of the background Minkowski metric into an effective one $ g_{\mu\nu},$ following the fundamental ideas of General Relativity;}
\item{The effective metric $ g_{\mu\nu}$ does not have a dynamics by its own but inherits the dynamics of the fundamental fields $ \Psi_{g}$ and $\Omega_{g}.$ }
\end{enumerate}

\section{The gravitational field}

In the present paper we restrict our analysis to the simplest case
in which there is only one g-neutrino $ \Psi_{g}.$  In a subsequent work
we will deal with the general case concerning also $ \Omega_{g}.$

Hence, the only contribution to the effective metric is given by $ \Delta_{\mu} $
and equation (\ref{27abril1}) reduces to
\begin{equation}
 h_{\mu\nu} =  \Delta_{\mu} \, \Delta_{\nu}.
 \label{7maio21}
 \end{equation}

A simple inspection shows that such tensor satisfy the properties
\begin{equation}
 h_{\mu\nu} \, \eta^{\mu\nu} = 0\,,
 \label{25abril1}
 \end{equation}
 \begin{equation}
 h_{\mu\nu} \, h_{\alpha\beta} \, \eta^{\nu\alpha} = 0.
 \label{25abril2}
 \end{equation}

 It then follows that it is possible to substitute in these formulas $ \eta_{\mu\nu}$ by $ g_{\mu\nu}.$ Furthermore, we note that due to the limitation to just one single spinor field, combining with the property that $ \Delta_{\mu}$ is a null vector imply that the determinant of $ g_{\mu\nu} $ is equal to the determinant of $ \eta_{\mu\nu}.$ We shall see that this limited framework contains some important metrics as for instance the case of a spherically symmetric and static field of a massive object.

The effect of the interaction of the g-neutrino with all kind of matter is what we call gravity. Following the path open by General Relativity this is realized by the substitution of the flat background metric into the curved one controlled by $\Psi_{g}.$ Besides, we must exhibit how matter of any sort affects such effective metric. Let us consider the example of the dynamics in which matter is represented by a massless scalar field $ \Phi$ interacting gravitationally. Its dynamics is given by the action:

\begin{equation}
S = \frac{1}{2}\int \, \sqrt{-g} \, g^{\mu\nu} \, \partial_{\mu} \Phi \, \partial_{\nu} \Phi\,.
\label{16abril3}
\end{equation}
This is the same as it occurs in General Relativity. The crucial difference appears when we look for the effect back of the matter on the spinor field. Let us calculate this explicitly. We set

\begin{equation}
\frac{\delta S}{\delta \overline{\Psi}_g} =   \frac{1}{2}\int \frac{\delta \sqrt{-g}}{\delta \overline{\Psi}_g} [\, g^{\mu\nu} \, \partial_{\mu} \Phi \, \partial_{\nu} \Phi] \,+\, \frac{1}{2}\int \sqrt{-g} \, \frac{\delta g^{\mu\nu}}{\delta \overline{\Psi}_g} \, [ \partial_{\mu} \Phi \, \partial_{\nu} \Phi ] = 0.
\label{27maio8}
\end{equation}

From the properties (\ref{25abril1}) -- (\ref{25abril2}) it follows that the first term vanishes identically. For the second term we must use other properties of Fierz-Pauli-Kofink identities to obtain
$$ \frac{\delta J}{\delta \overline{\Psi}_g} = \frac{J_{\mu} \, \gamma^{\mu} \, \Psi_g}{J}\,,$$ which implies
\begin{equation}
 \frac{\delta \Delta^{\nu}}{\delta \overline{\Psi}_g} = \left(\frac{g_{w}}{J^{2}}\right)^{1/4}  \left[ \gamma^{\nu} \, (1 - \gamma_{5}) \, \Psi_g - \frac{(J^{\nu} - I^{\nu}) \, J^{\mu}\,\gamma_{\mu} \,\Psi_g}{2 \, J^{2}} \right].
 \label{27maio2}
 \end{equation}

Then we obtain the dynamics of $ \Psi_g$ and of its source the scalar field $\Phi:$

 \begin{equation}
 i \, \gamma^{\mu} \, \nabla_{\mu} \Psi_g + 2 \kappa \,\left(\frac{g_{w}}{J^{2}}\right)^{1/4} \, (\Phi_{,\,\mu}\, \Delta^{\mu}) \, \Phi_{, \,\nu}\, \gamma^{\nu}\,(1 - \gamma_{5}) \,\Psi_{g}- \frac{\kappa}{J^{2}}\, (\Phi_{,\,\mu}\, \Delta^{\mu})^{2} \, (A + i B \gamma_{5})\,\Psi_{g}= 0,
 \label{25abril3}
 \end{equation}

\begin{equation}
\Box \Phi =  \frac{1}{\sqrt{-g}} \,\partial_{\mu} \left( \sqrt{-g} g^{\mu\nu} \, \partial_{\nu} \Phi \right) = 0.
\label{25abril5}
\end{equation}

By noting that the energy-momentum tensor of the scalar field is
$$ E_{\mu\nu} = \Phi_{, \, \mu} \, \Phi_{, \, \nu} - \frac{1}{2}  \Phi_{, \,\alpha} \, \Phi_{, \, \beta} \, g^{\alpha\beta} \, g_{\mu\nu}$$ we can rewrite (\ref{25abril3}) as
\begin{equation}
i \, \gamma^{\mu} \, \nabla_{\mu} \Psi_g + \kappa\, E_{\mu\nu} \, \left(\, \left(\frac{g_{w}}{J^{2}}\right)^{1/4} \,  \Delta^{\mu} \,  \gamma^{\nu}\,(1 - \gamma_{5}) \,  - \frac{1}{2J^{2}}\,  \Delta^{\mu} \, \Delta^{\nu}  \, (A + i B \gamma_{5})\, \right)\,\Psi_g = 0\,.
\label{27maio5}
\end{equation}
We remind the property $ \eta^{\mu\nu} \, \Delta_{\mu} = g^{\mu\nu} \, \Delta_{\mu}.$

As this case suggests it is not difficult to realize that the equation of motion of any kind of matter in gravitational interaction is the same as the one present in General Relativity like the scalar field in the above example.

 Note that the presence of the non-linear term $ A + i B \gamma_{5}$ appears also in other theories either in the realm of a complete theory of fields and elementary particles as suggested by Heisenberg in \cite {heisenberg} and \cite{heisenberg2} or the proposal of a dynamical model of elementary particles \cite{nambu}.

 For other forms of matter we can use the same approach, that is, we set for the matter dynamics similar to eq. ( \ref{16abril3})
\begin{equation}
S_{m} = \int \, \sqrt{- g} \, L_{m}
\label{27maio1}
\end{equation}
where the variation of the metric gives
$$  \delta \,  \sqrt{- g} \, L_{m} = \frac{\sqrt{- g}}{2} \, T_{\mu\nu} \, \delta g^{\mu\nu}. $$ Using the equation (\ref{27maio2}) it follows that the influence of any kind of matter on the dynamics of the g-neutrino  $ \Psi_{g}$ is given by
\begin{equation}
i \, \gamma^{\mu} \, \nabla_{\mu} \Psi_g \,+\, \kappa \, T_{\mu\nu} \, Q^{\mu\nu} \, \Psi_g = 0\,,
\label{27maio4}
\end{equation}
where
$$ Q^{\mu\nu} \equiv    \, \left(\,\frac{g_{w}}{J^{2}}\,\right)^{1/4} \,  \Delta^{\mu} \,  \gamma^{\nu}\,(1 - \gamma_{5}) \, - \frac{1}{2J^{2}} \,  \Delta^{\mu} \, \Delta^{\nu}  \, (A + i B \gamma_{5}).$$

\subsection{Spherically symmetric and static Gravitational field }

In order to understand the consequences of the present proposal let us examine some examples of the behaviour of the g-neutrino and the associated effective metric. In the present section we consider $\Psi_g$ without self-interaction (\ref{Heisenberg-interaction}).  We start by writing the Minkowski background (where the g-neutrino lives) in the $ (t, r, \theta, \varphi)$ coordinate system
\begin{equation}
ds^2 = dt^2 - dr^2 - r^2 \, d\theta^2 - r^{2} \, sin^2\theta \, d\varphi^2
\label{20abril 1}
\end{equation}
The corresponding $ \gamma_{\mu} $ in terms of the Euclidean ones $ \tilde{\gamma}_{\mu}$ are
\begin{align*}
\gamma_{0} &= \tilde{\gamma}_{0}\,, \\
\gamma_{1} &= \tilde{\gamma}_{1}\,, \\
\gamma_{2} &= r \, \tilde{\gamma}_{2}\,,\\
\gamma_{3} &= r \, sin\theta \,  \tilde{\gamma}_{3}\,.
\end{align*}

In the Appendix B we exhibit the convention for the constant $\tilde{\gamma}$\rq s. The non-identically zero Fock-Ivanenko coefficients are
\begin{align*}
 \Gamma_{2}^{FI} &= - \, \frac{1}{2} \, \tilde{\gamma}_{1} \tilde{\gamma}_{2}\,, \\[1.5ex]
\Gamma_{3}^{FI} &= - \, \frac{1}{2} \, cos \theta \, \tilde{\gamma}_{2} \tilde{\gamma}_{3}  -  \frac{1}{2} \, sin \theta \, \tilde{\gamma}_{1} \tilde{\gamma}_{3}\,.
\end{align*}

For the extra term, we have $ U_{\mu}$ from eq. (\ref{5abril3})  with $ H = \ln \sqrt{r \, \sin\theta}.$ We set for the field $$\Psi_{g}= f(r) \, B(\theta) \, \Psi^{0}\,,$$ where $ \Psi^{0} $ is a constant spinor. The dynamics provided by the Dirac\rq s equation  takes the form
\begin{equation}
\left[ \tilde{\gamma}_{1} \, \left( \frac{f^{'}}{f} + \frac{1}{2 r}\right)  + \tilde{\gamma}_{2} \, \frac{1}{rB} \, \frac{dB}{ d\theta} \right]\, \Psi^{0} = 0\,,
\label{23abril1}
\end{equation}
with $ f^{'} = df/dr.$ Thus the field depends only on coordinate $r$ that is
\begin{equation}
\Psi_g = \frac{1}{\sqrt{r}} \, \Psi^{0}.
\label{20abril 3}
\end{equation}

Let us set for the constant spinor the decomposition in terms of bi-spinors and write
$$ \Psi^{0}= \left(\begin{array}{l} \phi\\
\eta\\
\end{array}
\right).$$

We search for a particular solution such that
\begin{align}
(\phi^{+} - \eta^{+}) \, \sigma_{2} \, (\phi -\eta) &= 0
\label{7maio22} \\[1ex]
 (\phi^{+} - \eta^{+}) \, \sigma_{3} \, (\phi -\eta) &= 0
 \label{7maio23}
 \end{align}
where $\sigma_{i}$ are Pauli matrices.

Then it follows that $ \Delta_{2}= \Delta_{3} = 0.$
Using these results into the expression of $ \Delta_{\mu}$ we obtain $ \Delta_{0} = \textcolor{blue}{-}\Delta_{1} = g_{w}^{1/4} \, c_{0}/ \sqrt{r}$ where $c_{0}$ is a constant.
Thus it follows for the expression of the effective gravitational metric (\ref{9maio1}) the form

\begin{equation}
ds^2 = \left( 1 - \frac{r_{H}}{r} \right) \,dt^2 - \left(1 + \frac{r_{H}}{r} \right) \, dr^2 \textcolor{blue}{+} \frac{2 \, r_{H}}{r} \, dt \, dr - r^2 \, d\theta^2 - r^{2} \, sin^2\theta \, d\varphi^2\,,
\label{20abril5}
\end{equation}
where $ r_{H}$ is a constant. Making a coordinate change in order to eliminate the cross-term we set  for the new time $T$ the relation
$$ dt = dT \textcolor{blue}{-} \frac{r_{H}/r}{1 - r_{H}/r} \,dr $$
to obtain the final expression

\begin{equation}
ds^2 = \left( 1 - \frac{r_{H}}{r} \right) \,dT^2 - \frac{1}{ (1 - r_{H} /r)} \, dr^2 - r^2 \, d\theta^2 - r^{2} \, sin^2\theta \, d\varphi^2\,,
\label{20abril 6}
\end{equation}
which coincides with the solution of General Relativity. This is a very unexpected result indeed.

\subsection{A new gravitational black hole}

Let us consider the case in which the g-neutrino self-interact via the potential (\ref{Heisenberg-interaction}). The Dirac-Heisenberg\rq s equation is
\begin{equation}
 i\gamma^{\mu}
\nabla_{\mu} \, \Psi_g \,-\, 2 s \, (A + iB \gamma_{5}) \, \Psi_g = 0
\label{domingo2}
\end{equation}

Writting $\Psi(r) = f(r) \,\Psi^0$ and setting $ \sigma_{1} \phi = \phi,$ and $\sigma_{1} \eta = \eta,$
that is

\begin{align*}
\phi &= \left(\begin{array}{l} \alpha\\
\alpha\\
\end{array}
\right)\,, \\[1.5ex]
\eta &= \left(\begin{array}{l} \beta\\
\beta\\
\end{array}
\right)\,,
\end{align*}
the unique non-identically null currents are $ J^{\mu}, I^{\mu}$ for $ \mu = 0$ or $\mu = 1,$. Thus (\ref{domingo2}) take the form
\begin{equation}
	i \,  \, \left( f^{'} + \frac{f}{2 r}\right) \, \beta \, +\, f f^{*} f  \, \left( M \, \alpha - N \, \beta\right)  = 0
	\label{23abril1}
\end{equation}

\begin{equation}
	i \,  \, \left( f^{'} + \frac{f}{2 r}\right) \, \alpha \, -\, f f^{*} f  \, \left( M \, \beta - N \, \alpha \right)  = 0
	\label{12junho1}
\end{equation}
where $ M = 4 \, s \, (\alpha \, \alpha^{*} - \beta \, \beta^{*})$ and $N =  \, 4 \, \,s (\alpha^{*} \, \beta - \beta^{*} \, \alpha)$ with $ f^{'} = df/dr.$

Then it follows
\begin{equation}
	f^{'} + \frac{f}{2 r} -  \lambda \, f^{3} = 0
	\label{3junho2}
\end{equation}
where $ \lambda$ is a constant related to the constant spinor $ \Psi^{0}.$ The solution for real $\lambda$ \cite{comment} is given by
\begin{equation}
	f = \frac{1}{\sqrt{a_{0} \, r - 2 \,  \lambda \, r \log r}}
\end{equation}
\label{3junho5}

Using these results into equation (\ref{9maio1}) the effective gravitational metric takes the form

\begin{equation}
	ds^2 = \left( 1 - \frac{1}{S} \right) \,dt^2 - \left(1 + \frac{1}{S}   \right) \, dr^2 +  \frac{2}{S} \, dt \, dr - r^2 \, d\theta^2 - r^{2} \, sin^2\theta \, d\varphi^2\,,
	\label{20abril5}
\end{equation}
where $ S = r \, ( a_{0} - 2 \, \lambda \, \log r).$  Note that when the interaction vanishes $ ( \lambda = 0)$ the gravitational metric coincides with the Schwarzschild metric, which led to set $ a_{0} = 1/ 2 \kappa m.$

Making a coordinate change in order to eliminate the cross-term we set  for the new time $T$ the relation
$$ dt = dT - \frac{1}{S - 1} \,dr $$
to obtain the final expression

\begin{equation}
	ds^2 = \left( 1 - \frac{1}{S} \right) \,dT^2 - \left(1 - \frac{1}{S}\right)^{-1} \, dr^2 - r^2 \, d\theta^2 - r^{2} \, sin^2\theta \, d\varphi^2.
	\label{20abril 6}
\end{equation}
The horizon occurs for values of coordinate $ r = R_{H}$ given by
$$ \log \, R_{H} = \frac{1}{2 \lambda} \left[ \frac{1}{r_{H}} - \frac{1}{R_{H}} \right]$$
where $ r_{H}= 2 \kappa m $ is the Schwarzschild horizon.

A solution of this is given by
$$ R_{H} = - \, \frac{1}{2\lambda} \, \frac{1}{W(z)} $$
where $$ z = - \frac{1}{2 \, \lambda} \, e^{- \,1/2\lambda r_{H}}.$$
$W(z)$ is Lambert function defined by
$$ z = W(z) \, e^{W(z)}. $$
We can then compare the values of the horizon $ R_{H}$ of STG and $r_{H}$ for GR. Making an expansion of Lambert function to first order we can approximate
$$ R_{H} \approx 1 + \frac{1}{2 \, \lambda \, r_{H}}.$$  It then follows that, for positive values of $ \lambda$  we obtain
\begin{itemize}
	\item{ For $ 0 < r_{H} < 1 + \sqrt{1 + 2/\lambda}$ then $R_{H} > r_{H};$}
	\item{$ r_{H} > 1 + \sqrt{1 + 2/\lambda}$ then $R_{H} < r_{H}.$}
\end{itemize}

\section{Cosmological framework}

In this section we start the program of generating a new cosmology in the spinor theory of gravity (STG). Here we limit ourselves to describe an expanding universe without matter. The cosmological solution of an empty universe in General Relativity is the Kasner metric. We shall prove that there is an empty universe in STG that represents an isotropic cosmos.

Consider the Minkowski background

\begin{equation}
ds^2 = dt^2 - dx^2 - dy^2 - dz^2
\label{20maio5}
\end{equation}
The internal connection (\ref{5abril5}) reduces to
$$ \Gamma_{\mu} = \frac{\varepsilon}{4} \,  \gamma_{\mu} \, \gamma^{\alpha} \, H_{\alpha}.$$
The simplest solution of the equation $ \Box H = 0$ provides $ H = m \, t$. The equation for $\Psi_g$ becomes
\begin{equation}
\gamma^{\alpha} \, \partial_{\alpha}\Psi_{g}- \lambda \, \gamma^{0} \,\Psi_{g}= 0
\label{2omaio6}
\end{equation}
where $ \lambda = \varepsilon \,m.$ The solution is immediate :
\begin{equation}
\Psi_g = e^{\lambda \, t} \, \Psi^{0}
\label{20maio7}
\end{equation}
where  $ \Psi^{0} $ is the constant spinor
$$ \Psi^{0}= \left(\begin{array}{l} \phi\\
\eta\\
\end{array}
\right).$$
The arbitrariness on  $ \Psi^{0} $ allow us to set
$ (\phi^{+} \, \sigma_{k} - \eta^{+} \, \sigma_{k}) \, (\eta - \phi)$ to have the same value for $ k = (1, 2, 3).$ It then follows
$ \Delta_{1} = \Delta_{2} = \Delta_{3} = \Delta_{0}/ \sqrt{3}.$ Let us then call $ \Delta_{0} = \Delta$ which has the value
$$ \Delta \propto  \, e^{\lambda \, t}$$
and which yields the form of the effective metric, that is

\begin{align*}
ds^2 = (1 - \Delta^2) \, \, dt^2 -   (1 + \frac{\Delta^{2}}{3}) \, ( dx^2 + dy^2 + dz^2) - \frac{2}{\sqrt{3}} \, \Delta^{2} \, dt \, ( dx + dy + dz) - \frac{2}{3} \, \Delta^{2} \, ( dx \, dy + dx \, dz + dy \, dz). \label{20maio8}
\end{align*}

Making a transformation of coordinates from $ (t, x, y, z)$ to $ ( t, u, v, q)$ we set

$$  dx = du + dv + dq + m(t) dt.$$
$$  dy = du + dv + dq + m(t) dt.$$
$$  dz = du + dv + dq + m(t) dt.$$
and choosing $ m(t)$ to eliminate cross-terms of the form $ g_{0i}$ we obtain

\begin{equation}
ds^2 = ( 1 + \Delta^{2} )^{-1} dt^2 - 3 \, (1 + \Delta^{2}) \left[ du^2 + dv^2 + dq^2 + 2 \, ( du \, dv + du \, dq + dv \, dq) \right]\,.
\label{20maio9}
\end{equation}
Redefining the spatial coordinates as
$$ dX = \alpha \, \left( p_{1} \, du + p_{2} \, dv + p_{3} \, dq \right) $$
$$ dY = \alpha \, \left( p_{2} \, du + p_{3} \, dv + p_{1} \, dq \right)$$
$$ dZ = \alpha \, \left(p_{3} \, du + p_{1} \, dv + p_{2} \, dq \right)$$
  where the constants $ p_{i} $ satisfy the two conditions
  $$ p_{1} + p_{2} + p_{3} = \frac{3}{\alpha} $$
  $$ p_{1}^{2} + p_{2}^{2} + p_{3}^{2} = \frac{3}{\alpha^2} $$
  we obtain
  \begin{equation}
  ds^2 = ( 1 + \Delta^2)^{-1}  \, dt^2 -  \, (1 + \Delta^{2} ) \left( dX^2 + dY^2 + dZ^2 \right).
  \label{22maio1}
  \end{equation}

Making a new definition of time by setting $$ 1 + \Delta^2 = 1 + \exp(2 \sigma t) = \tanh^2 T $$
gives the cosmological solution for the effective metric for the STG in Gaussian global time
  \begin{equation}
  ds^2 = dT^2 - \tanh^2 T \, \left(dX^2 - dY^2 - dZ^2 \right)\,,
  \label{22maio2}
  \end{equation}

\section{Conclusion}

Since the success of the unified description of the weak and electromagnetic interactions the gauge scenario became accepted as the natural road to undertake the unification of all forces in nature. The present work goes in an opposite direction and follow an old idea that weak interaction and gravity have an intimate relationship. Although these forces are described by very distinct languages the possibility to obtain a unique framework through the uses of the modern understanding of the meaning of an effective metric opens a new way to accomplish this task and overcomes this difficulty.

The most remarkable aspect of GR rests on the idea that gravity concerns to the modification of the geometry of space-time. This is precisely the point of start of the spinor theory of gravity (STG). Indeed, STG is based on the fundamental principle of GR which states that any gravitational phenomena can be described in terms of an effective modification of the Riemannian geometry of space-time. The most deep distinction between GR and STG concern the crucial hypothesis made by GR which imposes a dynamics of its own to the metric $ g_{\mu\nu}.$

In an alternative proposal STG states the hypothesis that such geometry is not a dynamical field but instead it is the consequence of a more deep structure that we attribute to the role of the spinorial fields called \textit{g-neutrinos} $ \Psi_{g}$ and $\Omega_{g}. $ In this vein the effective metric inherits the dynamics of these two fields. On its basic motivation STG deals with null-currents typical of Fermi interaction.

We describe in this way how matter and energy of any sort affects the gravitational field and conversely how they are affected by gravity. As a consequence of this connection with Fermi\rq s interaction one should think if gravitational processes could also violate parity \cite{parity}. We will consider this question in a subsequent work.

In addition, three solutions of STG are presented  -- two concerning to static and spherically symmetrical configurations (with and without self-interaction, respectively) and another one applied to cosmology\footnote{After have had submitted the present paper, we communicated in \cite{gravwaves} a possible direction in order to define gravitational waves in STG.}. 

As an unexpected result, the effective metric in the first case coincides with the Schwarzschild solution of GR. In the second case, when the g-neutrino self-interact via Heisenberg's potential, a new kind of black-hole emerges. This new type of static and spherically symmetric gravitational field is setting the first possibility of falsify the STG. An analysis of this solution is in progress. 

Finally, the second class of testability of the STG comes from the program of cosmological frameworks initiated in the Section VIII. An empty universe produces a geometry equivalent to the case of GR generated by a perfect fluid endowed with the equation of state of stiff matter: $ p = \varrho.$

\section{Appendix: mathematical compendium}
\subsection{The formula of the determinant}

In order to evaluate the determinant of the metric under the decomposition
$$ g_{\mu\nu} = \eta_{\mu\nu} + h_{\mu\nu}$$
we can use the formula of the determinant provided by Cayley-Hamilton theorem that yields the expression of the matrix $ T$ as:

$$
\det\textbf{T} = -\frac{1}{4}\left[{\rm Tr} (\textbf{T}^{4}) - \frac{4}{3} \,{\rm Tr} (\textbf{T}) \, {\rm Tr} (\textbf{T}^{3}) - \frac{1}{2} \, \left({\rm Tr} (\textbf{T}^{2})\right)^{2} + \left({\rm Tr} (\textbf{T})\right)^{2} \, {\rm Tr} (\textbf{T}^{2}) - \frac{1}{6} \, \left( {\rm Tr} (\textbf{T}) \right)^{4}\right].
$$

We note that in the case of a unique current the determinant of the effective metric $ g_{\mu\nu}$ is equal to the determinant of the background metric $\eta_{\mu\nu}$ once the metric $ g_{\mu\nu}$ is constructed with the single weak current that is a null vector.

\subsection{The convention for the gamma\rq s}

For completeness we exhibit our notation of the constants gamma\rq s:

\begin{align*}
    \widetilde{\gamma}_o =  \left(\begin{array}{cc}
    I_2 & 0 \\
    0 & -I_2
    \end{array}\right)\,,
    \qquad
    \widetilde{\gamma}_k =  \left(\begin{array}{cc}
    0 & \sigma_k \\
    -\sigma_k & 0
    \end{array}\right)\,
    \qquad
    \widetilde{\gamma}_5 =  \left(\begin{array}{cc}
    0 & I_{2} \\
    I_{2} & 0
    \end{array}\right).
    \end{align*}
\begin{align*}
\sigma_1 =  \left(\begin{array}{cc}
    0 & 1 \\
    1 & 0
    \end{array}\right)\,,
    \qquad
\sigma_2 =  \left(\begin{array}{cc}
    \,0\, & \,-i\, \\
    \,i\, & \,0\,
    \end{array}\right)\,,
    \qquad
\sigma_3 =  \left(\begin{array}{cc}
    1 & 0 \\
    0 & -1
    \end{array}\right)\,.
\end{align*}

\section{acknowledgements} M Novello thanks a fellowship from CNPq and FAPERJ. We  thank Prof. Roberto Onofrio for calling our attention to their interesting papers  and arguments.


\begin{thebibliography}{100}
\bibitem{comment}
The general case of equation (\ref{3junho2}) can be written $ \lambda = a + i b.$ Then the solution for the metric is
$$ S \equiv \frac{1}{f^2} = r \left[\, \left(\,\frac{1}{r_{H}}\,\right)^2 \,-\, 4 \, \frac{a}{r_{H}} \, \log{r} \,+\, 4 (a^2 - b^2) \log^{2} r\right]^{1/2}\,. $$
\end{thebibliography}
 \end{document}